\documentclass[journal=aamick,manuscript=article]{achemso}

\usepackage{url}
\usepackage{graphicx}
\usepackage{dcolumn}
\usepackage{bm}
\usepackage{amssymb,amsmath}
\SectionNumbersOn

\author{Andreas Pedersen}
\affiliation{Integrated Systems Laboratory, ETH Zurich, 8092 Zurich, Switzerland}
\email{andped10@gmail.com}
\author{Mathieu Luisier}
\affiliation{Integrated Systems Laboratory, ETH Zurich, 8092 Zurich, Switzerland}

\title{Lithiation of Tin Oxide: A Computational Study}

\begin{document}

\begin{abstract}
We suggest 
that the lithiation of pristine SnO forms a layered Li$_\text{X}$O structure 
while 
the expelled tin atoms agglomerate into 'surface' planes separating the Li$_\text{X}$O layers. The proposed lithiation model widely differs from the common assumption that tin segregates into nano-clusters embedded in the lithia matrix. 
With this model we are able to account for the various tin bonds that are seen experimentally and explain the three volume expansion phases that occur when SnO undergoes lithiation: 
(i) at low concentrations Li behaves as an intercalated species inducing small volume increases;
(ii) for intermediate concentrations SnO transforms into lithia causing a large expansion; 
(iii) finally, as the Li concentration further increases a saturation of the lithia takes place until a layered Li$_2$O is formed. A moderate volume expansion results from this last process.   
We also report a 'zipper' nucleation mechanism that could provide the seed for the transformation from tin oxide to lithium oxide.

\end{abstract}


\section{Introduction}
\label{sec:Introduction}
\begin{figure}
\centering
\includegraphics[width=.5\textwidth]{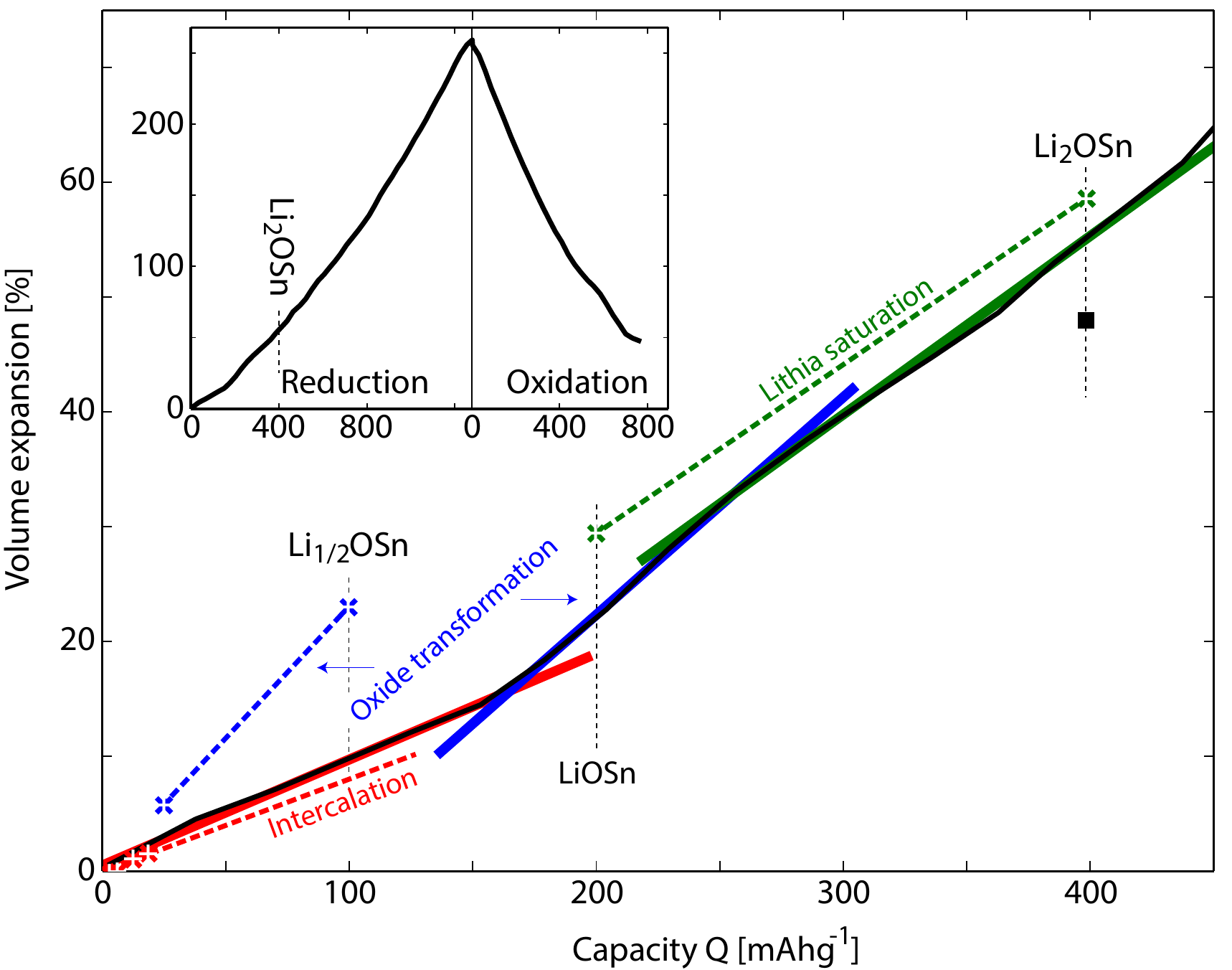}
\caption{Relative volume expansion of Li$_{\text{X}}$OSn with respect to a SnO sample during the first lithiation cycle. The reported curve only represents the initial part of the full lithiation/delithiation cycle that is shown in the inset. The black line refers to the experimental results of Reference~\citenum{Ebner:2013bj}. The solid black square, which is also taken from Reference~\citenum{Ebner:2013bj}, represents the volume expansion for the case where Li$_2$O and Sn completely segregate.
Three regimes can be identified from the measured curve, which are indicated by the colored lines
(i) red, intercalation (ii) blue, oxide transformation and (iii) green, saturation of the lithia. The crosses are results from our 
%
simulations. 
They are colored according to the expansion regime they correspond to.  
The dashed lines connect computed data points belonging to the same regime.
}
\label{fig:expansion}
\end{figure}

Tin based compounds represent a promising category of materials to be used 
as
 the anode of lithium-ion batteries (LIBs)\cite{Obrovac:2007fw}. With an observed capacity of either 4.4 (Li$_{22}$Sn$_5$)~\cite{Boukamp:1981iz} or 4.25 (Li$_{17}$Sn$_4$)~\cite{Goward:2001iu} Li per Sn atom, they outperform the maximal theoretical energy density of graphite electrodes by a factor greater than two\cite{Park:2010kj}.
%
%
The interest in Sn-based systems was triggered by the work of Idota {\it et al.}~\cite{Idota:1997co} who applied an amorphous tin composite oxide as the anode material. Several other research groups have 
later
investigated similar compounds in the context of LIBs~\cite{Courtney:1997wl, Sandu:2004gl, Ebner:2013bj}. However, a commercially viable solution remains to be determined since Sn-based systems tend to suffer from an unacceptably fast fade in capacity. This happens in spite of the various attempts to optimize the chemical composition\cite{Kepler:1999bo, Beaulieu:2000vq, Zhang:2006ec} and structural features~\cite{Zhang:2008ef, Xu:2012ey, Wang:2012bq, Kravchyk:2013iy} of the anode, showing that a better understanding of the atomic structure of the active materials is required for improved designs.
A recent review by Park {\it et al.}\cite{Park:2010kj} lists the most relevant advances made in Sn-based electrodes and their main causes of degradation.
The large volumetric changes of Sn, SnO, and SnO$_2$ during lithiation and delithiation is one of the main deterioration mechanism and among the hardest technological challenges to overcome. An example of the expansion taking place when SnO is lithiated is displayed in Figure~\ref{fig:expansion}, which reports the volumetric change of Li$_\text{X}$OSn with respect to the initial SnO crystal as a function of the Li load level\cite{Ebner:2013bj}. At moderate Li concentrations, i.e. up to a ratio of two Li atoms per Sn atom, three distinct expansion regimes can be identified from this experimental data, as discussed later.

In one of the first studies addressing the structural changes occurring during the lithiation of Sn-based compounds (SnO, SnO$_2$, Li$_2$SnO$_3$, and SnSiO$_3$) Courtney and Dahn~\cite{Courtney:1997wl} suggested that the initial tin oxide transforms into 
an amorphous
lithia matrix, from which the Sn atoms are expelled before segregating into nano-clusters embedded within the forming matrix. This lithiation model relies on two phenomena. First, an irreversible process takes place during the first lithiation cycle, which can be attributed to the formation of a stable lithia matrix. Secondly, XRD measurements of the lithiated samples contain signatures of 
Sn-Sn and Sn-Li
bonds, potentially validating the proposed clustering of Sn atoms. 
In a later work the transformation towards a lithia was confirmed by Chouvin {\it et al.}\cite{Chouvin:2000kl} who supported
this
with results from M\"ossbauer spectroscopy. They revealed that the Sn atoms are reduced as the Li concentration level increases. However, their data also showed that the forming lithia 
does 
not correspond to its bulk counterpart since markers for various unknown bonds were present.
A similar conclusion was reached by Sandu {\it et al.}~\cite{Sandu:2004gl} who also applied M\"ossbauer spectroscopy. Their results agian showed signatures of unidentified 
Sn
bonds after the transformation into lithia had taken place; they named them 'exotic' bonds. Furthermore they  observed that a spectra consistent with SnO was present at low Li concentrations and therefore concluded that Li behaves as an intercalated species during the initial part of the lithiation process.
As the Li load level increases they
also
found that parts of the SnO matrix could still accommodate Li atoms as intercalated species, while other parts started transforming into lithia.

Another peculiarity of Sn-based systems was later reported by Zhong {\it et al.}~\cite{Zhong:2011hs} 
and further investigated by Nie {\it et al.}~\cite{Nie:2013fu}
In their experiments they saw that the transformation into lithia occurs along stripes when a nanowire of SnO$_2$ gets floated with lithium. Such a nucleation is very different from the usually assumed core-shell mechanism and remains to be understood. Although the crystalline structure of SnO and SnO$_2$ are significantly different it is worth investigating whether Li preferably arranges itself along stripes in other Sn-based oxides.
From the experimental results summarized above it appears that 
(i) three expansion regimes can be seen when SnO is lithiated\cite{Ebner:2013bj},
(ii) Li atoms behave as intercalated species at low concentrations,
(iii) 'exotic' Sn bonds manifest themselves after the first lithiation cycle, and
(iv) an abnormal nucleation mechanism occurs in SnO$_2$.
None of the existing lithiation models simultaneously captures all these phenomena and offers a consistent explanation for them. 
Moreover, the common models assuming a complete segregation of Sn and Li$_2$O underestimates the experimentally determined volume expansion by 7.5\%, clearly indicating that important effects are not taken into account.

To address these issues we report here results
from simulations
investigating the structural evolution  during the lithiation of pristine SnO. Starting from a SnO crystal, atomistic simulations are conducted to determine the volume expansion as the oxide transforms into lithia up until the point where a saturated Li$_2$O$^{\text{L}}$ is formed. 
The final structure is layered and different from its crystalline counterpart and is therefore marked with a superscript L. We also demonstrate that rather than segregating into nano-clusters the Sn atoms remain evenly distributed between the forming Li$_2$O$^{\text{L}}$ layers so that they can be considered as 'surface' planes on the 
appearing
lithia. This new model challenges the current understanding of SnO lithiation as being a process where a Li$_2$O matrix forms and the expelled Sn atoms segregate into nano-clusters.
The paper is organized as follows: we start by describing the computational approach. Simulation results are listed in Section~\ref{sec:Results}, which is divided into two parts, one dealing with tin oxide and the other with lithia. Section~\ref{sec:Discussion} discusses the obtained results and their consistency with experimental data. Finally, conclusions are drawn in Section~\ref{sec:Conclusions}.

\begin{figure}
\centering
\includegraphics[width=.5\textwidth]{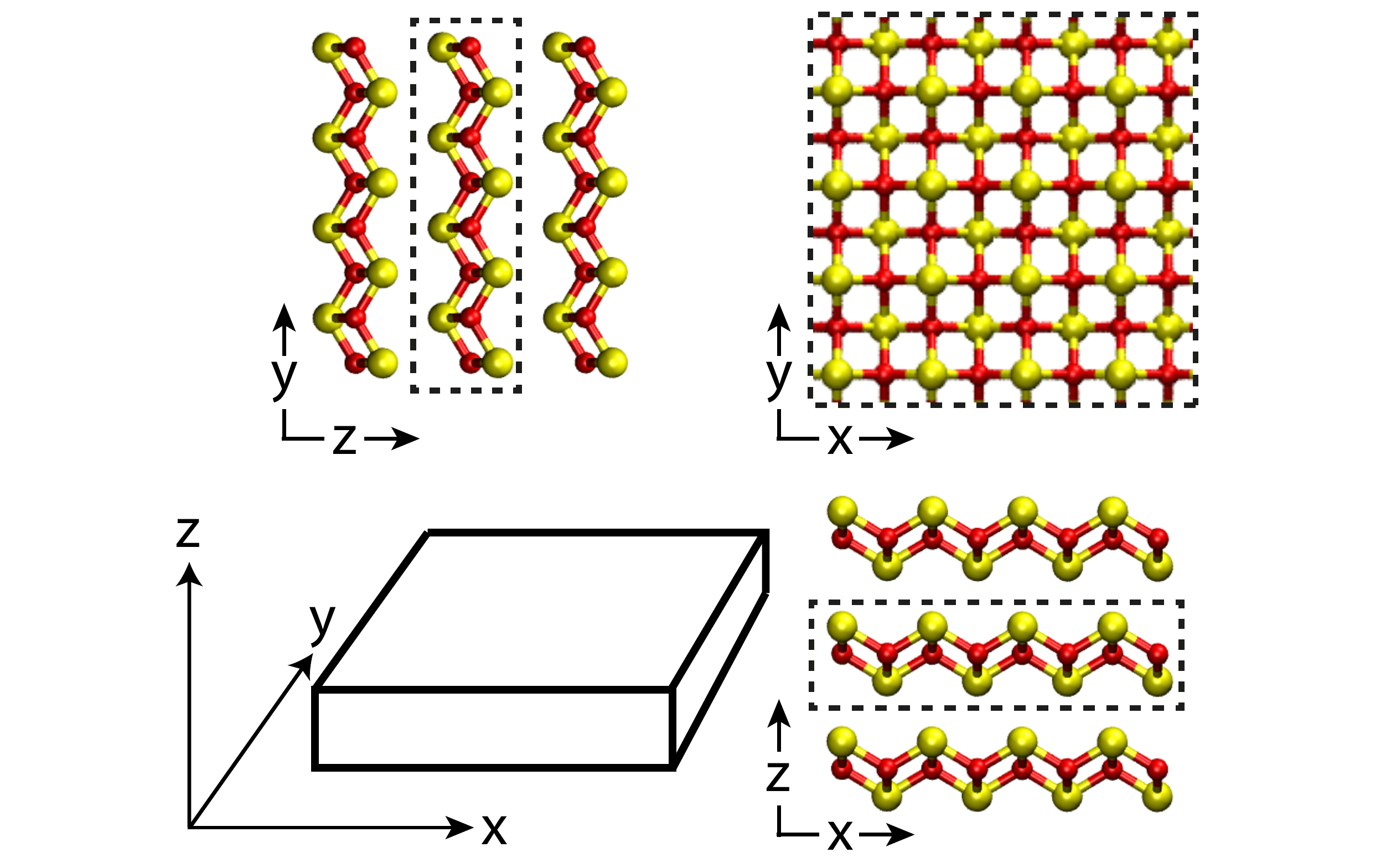}
\caption{
Pristine SnO (color online).  Yellow colored atoms represent tin and red oxygen. Only bonds shorter than 2.3~\AA\ are shown.
It is know that tin oxide is a layered and ordered structure~\cite{Wells:1984uu}. Here three such layers are shown and it appears that the (100) and (010) planes are identical. 
Each layer consists of two flat Sn planes and a slightly wiggling central plane of O atoms. Within each layer the Sn and O atoms connect to four neighbors of the opposite species.  The center most layer---enclosed in the dashed rectangular box---forms one of the two layers included in the simulated supercell.
}
\label{fig:SnO}
\end{figure}

\section{Computational Details}
\label{sec:computational}
Two types of supercells are utilized to determine the structural changes of SnO as it lithiates. In the first series of simulations addressing  low Li concentrations in SnO and the formation of the Li$_{1/2}$OSn crystal, the supercell consists of two layers, whereas it only contains a single layer for the calculations of LiOSn and
Li$_2$O$^{\text{L}}$Sn.
The definition of a layer is given in Figure~\ref{fig:SnO}. Both supercell types exhibit a total of 64 SnO units (64 Sn and 64 O atoms), while the number of Li atoms varies from 0 to 128.
%
 

The atomic interactions are accounted for 
by applying
the VASP~\cite{Kresse:1996vk} density functional theory (DFT) tool.  Core electrons are described by 
the projector augmented wave (PAW) method~\cite{Blochl:1994uk}.  All simulations are performed within the Perdew-Burke-Ernzerhof (PBE) parametrization of the generalized gradient approximation (GGA)~\cite{Perdew:1996ug} to the exchange-correlation functional. The KPOINT sampling is 2x2x2 and the plane-waves are truncated at 500~eV. The electronic degrees of freedom are considered as converged when the change in energy decreases below $1 \cdot 10^{-8}$~eV per self-consistent field (SCF) iteration. 
A Nose-Hover thermostat at either 1000K or 500K is applied in certain cases to enable transitions between nearby local minima. Finally, all structures are relaxed in terms of their ionic degrees of freedom 
(atomic nuclei)
and the size and shape of the supercell through the conjugate gradients method as implemented in VASP. In this case convergence is reached when the force decreases below $1 \cdot 10^{-3}$~eV/\AA. 

\section{Results}
\label{sec:Results}
\subsection{Tin oxide}
At low concentrations the inserted Li atoms have a preference for linear structures within the SnO matrix, i.e. they are favorably arranged as stripes. The lowest energy configurations result when the Li atoms reside in a single Sn-O-Sn layer and neighbor Li atoms belong to alternating Sn 'surfaces'. At this concentration level the Li atoms behave as intercalated species and are likely subject to diffusion within the plane spanned by a Sn-O-Sn layer. In the following two sections these findings are explained in greater details. 

\subsubsection{Pure oxide}
A pristine tin (II) oxide crystal is generated from a tetragonal PbO lattice in which the Pb atoms are replaced by Sn and relaxed~\cite{Wells:1984uu}. The resulting bulk phase of SnO has a distinct layered Sn-O-Sn structure in the Z-direction, which clearly appears in Figure~\ref{fig:SnO} that shows three such layers. Each of them can be divided into three planes: a central plane of 
four-fold coordinated
O atoms 
in tetrahedral sites
%
is sandwiched between two 'surface' planes of four-fold coordinated Sn atoms that reside at the apex of a square pyramid. 
The bond length between Sn and O is 
2.3~\AA,
which agrees reasonably well with the experimental determined value of 2.21~\AA\cite{MooreJr:1941bj} 
and a distance of 2.3~\AA\ is used as an the upper value to determine whether two atoms are connected by a bond. 
All the atoms have four nearest-neighbors and the resulting supercell is a tetragonal box with identical (100) and (010) planes. The distance between two O planes is 
4.9~\AA\ and 
the footprint per SnO unit in the (001) plane is 
7.4~\AA $^2$
 for a total volume of
36.7~\AA $^3$
per unit. 
Within the SnO structure the distance between the upper and lowermost Sn 'surface' planes of two adjacent layers is 
2.6~\AA,
which exceeds the height of the layer itself. 

In the limit of a single Li atom added to the SnO matrix, the site preference is within one of the Sn planes. The lowest energy configuration is found when the Li atom occupies the same site as a Sn atom belonging to the opposite 'surface' separated by the central O plane. 
In other words, the O plane acts as a mirror between the newly introduced Li atom and an existing Sn element.
In this configuration the binding energy for the atom is -250~meV and it is not thermodynamically favorable compared to the cohesive energy of a Li atom in a bulk environment. 
When located in this site the Li atom does not cause any significant disturbances of the SnO matrix and the bond length between neighbor Sn and O atoms remains equal to  
2.3~\AA.
The distance from Li to O is 
2.2~\AA.
The separation between Li and the closest Sn atoms in the same plane is
2.8~\AA,
while the distance to its mirror Sn atom residing in the opposite plane is
2.5~\AA.
These facts combined with the relative large spacing between the SnO layers clearly show that at low concentrations the Li atoms can be considered as intercalated species within the SnO matrix. Furthermore, at this concentration level, Li atoms are likely subject to 2D diffusion within the (001) plane.  

\subsubsection{Lithium at low concentrations}
\label{sec:intercalation}
\begin{figure}
\centering
\includegraphics[width=.75\textwidth]{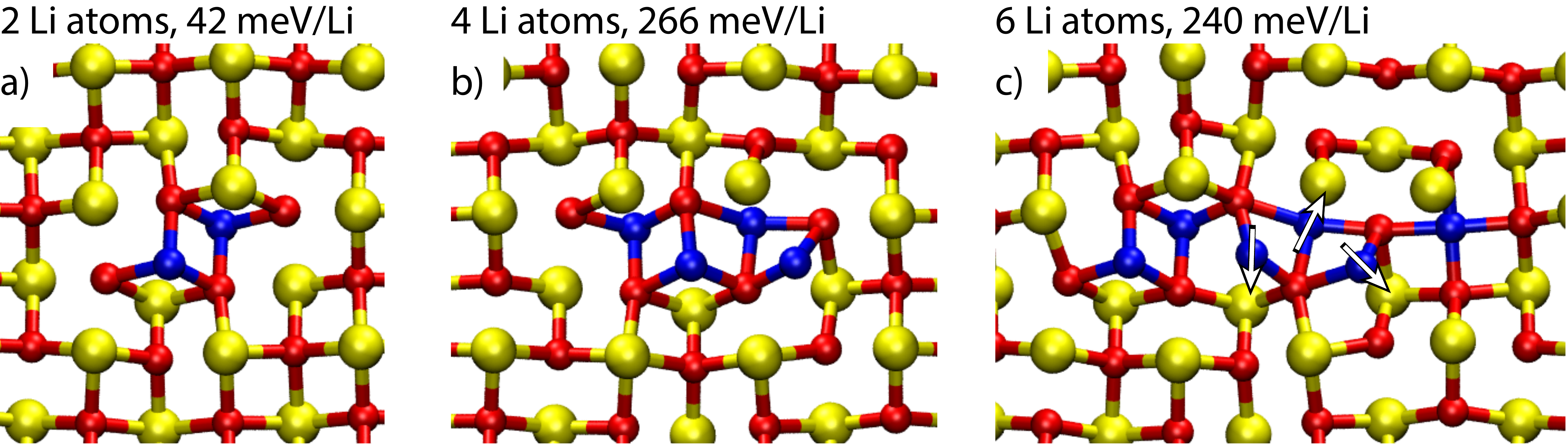}
\caption{Intercalation of Li in pristine SnO  (color online). The same color code as in Figure~\ref{fig:SnO} is used with the addition of blue for lithium.
a) A Li dimer with an atom in the Sn planes above and below the central O plane is the lowest energy configuration for two Li atoms.
b) Four Li atoms form a cluster with two atoms in the planes above and below the central O plane. This results in the strongest binding energy per Li atom among all the samples where Li is considered as an intercalated species. 
c) Nucleation and 'zipper' mechanism for the transformation from tin oxide to lithia. The arrows mark the direction along which Li atoms should move to form a stripe of lithia. It is important to note that the Sn atoms evenly move to the 'surface' sites situated above and below the forming LiO.}
\label{fig:intercalated}
\end{figure}

To investigate how lithiation affects the SnO matrix at low concentrations, numerous atomic configurations are generated where the load level increases from two to
eight
Li atoms.
In the search for the lowest energy structure, configurations are sampled with the additional Li atoms arranged to 
(i) reside within the same or neighbor SnO layers, 
(ii) be nearest-neighbors or separated, and 
(iii) ordered in different patterns.
Structures with four or more Li atoms are generated by embedding Li into a low energy configuration with less Li atoms. This technique allows for a more rapid convergence towards the lowest energy structures. 
To relax the stress caused by the inserted Li atoms, the heat treatment mentioned in Section \ref{sec:computational} is applied. The determined lowest energy structures for three of the investigated load levels are shown in  
Figure~\ref{fig:intercalated}. A complete list of the tested configurations can be found in the 
supporting information.

When the second Li atom is inserted into the SnO matrix the ratio between Li and Sn becomes 2:64.  The two Li atoms then have the possibility to arrange themselves in a dimer configuration.
A weak binding of 42~meV/atom is 
achieved 
%
when
they form a symmetric structure and reside in opposite Sn planes.  
By comparing to the structure with a single Li atom in the SnO matrix it appears that 292 meV is gained per atom and an energy barrier of 584~meV
has to be
 surpassed for the dimer to break.
In this configuration the distance to the neighbor O atoms is reduced to 
1.9~\AA\ instead of 2.2~\AA\ as for the single Li atom. 
The separation between the Li atoms is 
2.6~\AA,
0.1~\AA\
shorter than the  
distance between two O atoms within the same plane.  
Finally, the bond length to the closest Sn atoms within the same plane
slightly increases from 2.8 to 2.9~\AA, while the distance to the ones in the opposite plane within the same layer changes from 2.5 to 2.7~\AA.
When the dimer forms the surrounding SnO environment is somewhat disturbed and a few bonds connecting Sn and O atoms are stretched beyond the
defined bond
value, as can be seen in Figure~\ref{fig:intercalated}.    
The volume increase is insignificant as compared to the pristine SnO.

If an additional pair of atoms is inserted, a total of four Li atoms is embedded in the SnO matrix and a 4:64 ratio is reached. At this load level the lowest energy configuration
is
 a cluster with two atoms in each of the Sn 'surfaces' as shown in Figure~\ref{fig:intercalated} (b). The corresponding volume expansion is 1.03\%. The binding energy now increases to 266~meV/atom, which is the highest value per Li atom in any of the inspected low concentration structures.  The separation between the central pair of Li atoms is 
2.7~\AA,
whereas the distance between the rightmost pair is 
2.6~\AA, and  
2.4~\AA\
for the leftmost one.
The closest Sn atoms are situated in the opposite plane at a distance ranging from 
2.7 to 2.9~\AA.
The shortest distance from Li to a neighbor O atom is in all cases close to 
1.9~\AA.
It is noteworthy that the Li-O bond length has a very narrow spread. In fact, this value corresponds to the Li-O bond length in the fully saturated Li$_2$O$^\text{L}$ layer, which is described in a later section and stands out as a strongly desired structural feature. 
The four-atom cluster lacks symmetry because it seems impossible to conform to the surrounding SnO matrix (bond length of 2.3~\AA), while preserving 
%
the bonds to the O atoms.
%

When six Li atoms are present in the oxide matrix a ratio of 6:64 is reached. The relaxation of the atomic structure now leads to a replacement of one Sn atom in the oxide matrix by one of the inserted Li atoms. As a consequence the Sn atom is pushed outwards and away from the central plane when the Li atom situated in opposite plane enters the O plane. 
This transition spontaneously occurs indicating that no energy barrier needs to be surpassed for lithia to nucleate.  
After the transformation the binding energy per Li atom slightly reduces as compared to the configuration containing a cluster of four Li atoms and is equal to 240~meV/atom. The volume expansion reaches 1.55\%. In the relaxed structure the distances between neighbor Li atoms are 
2.5~\AA\
in three of the five cases, 
whereas for the last two the distance increases to
2.9 and 3.1~\AA.
The latter length corresponds to the case where a Li atom is within the central O plane.
%
Again the shortest bond 
from Li to a neighbor O atom is 
1.9~\AA\
and the closets Sn atoms are within 
a
range of 
2.7
to
3.0~\AA.

Under the assumption that linear arrangements of Li atoms remain at higher concentrations, the observed nucleation mechanism will act as a 'zipper' that joins the Li atoms and form a stripe of lithium oxide. The arrows in Figure~\ref{fig:intercalated} (c) indicate the directions of propagation for the Li atoms as they enter into the O plane. It is also important to note that during this transformation the Sn atoms will evenly be expelled above and below the emerging stripe of lithia. 

\begin{figure}
\centering
\includegraphics[width=.5\textwidth]{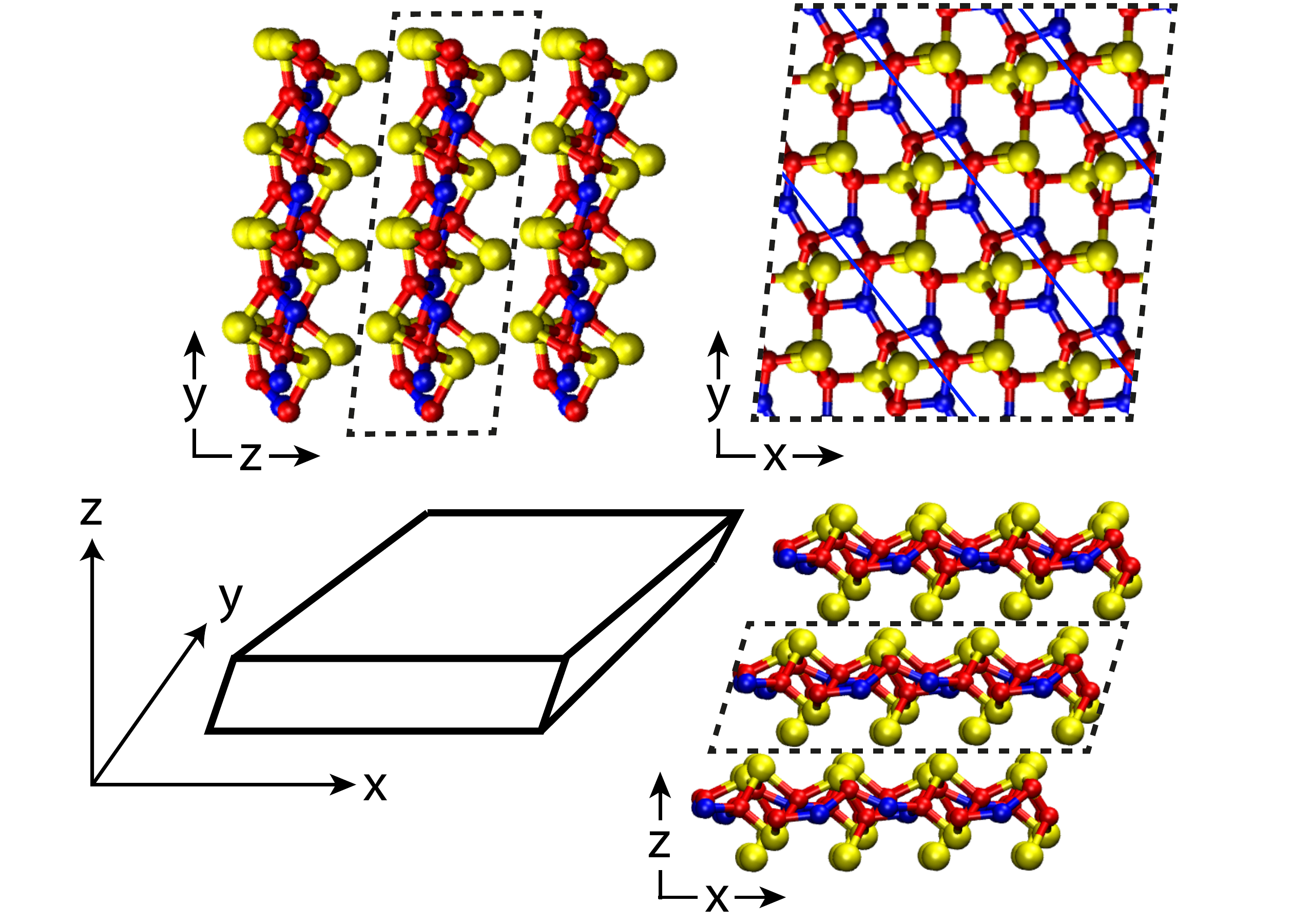}
\caption{Relaxed Li$_{1/2}$OSn sample (color online). 
The same color code as used before. The dashed box encloses the simulated supercell. The blue lines are imposed to underline the striped structure of the Li atoms within the transformed oxide.
As Li enters sites within the central O plane Sn atoms are pushed out. This causes a significant change in the bond configuration. For all the Sn atoms situated above the central O plane the coordination is reduced to three. Half of the Sn atoms below the central plane are now only bonded to two neighbor O atoms. 
The layered structure of the initial tin oxide remains, however, the supercell is skewed in the (001) plane and along the Z-direction.  
}
\label{fig:LiO_half_Sn}
\end{figure}

\subsection{Lithia} 
At moderate Li concentrations a sudden acceleration of the volume expansion occurs, which is explained by the transition from SnO to LiO. We will not go into the details of this process, but rather focus on the resulting Li$_{1/2}$OSn, LiOSn, and Li$_2$O$^\text{L}$ configurations.  Since these structures are fully transformed oxides without any traces of intercalated Li, the computed volumes are upper bounds for the corresponding capacities, which is also apparent in Figure~\ref{fig:expansion}. 
Despite the misaligned 'volume expansion vs capacity' these configurations are investigated to determine whether the layered structure of the initial SnO persists and if the expelled Sn atoms are still evenly distributed above and below the central plane.    

\subsubsection{Li$_{1/2}$OSn} 
\label{sec:LiO_half_Sn}

%
To gain insights into the structural changes taking place at moderate Li concentrations an atomic configuration with a  1:2 ratio between Li and Sn is constructed.  The structure is made from a pristine SnO crystal where lines of Li atoms are inserted into the Sn planes above and below the central O plane. Based on the findings at low Li concentrations in Section~\ref{sec:intercalation} the Li atoms are placed at mirrored Sn sites. Every second row is left empty.  
Of all the simulated samples Li$_{1/2}$OSn is the one that imposed the greatest challenge with several calculations resulting in layered, but disordered structures. 
However, it turns out that the structure with the lowest energy exhibits a layered and ordered arrangement, 
as shown in  Figure~\ref{fig:LiO_half_Sn}. It should be noted that the Li atoms form linear chains, which resemble the stripes experimentally seen during the lithiation of 
SnO$_2$~\cite{Zhong:2011hs}. 
%
Although the materials are different (SnO vs. SnO$_2$) linear chains of Li also appear in SnO. This could be a typical feature of Sn-based oxides.

Compared to the initial SnO crystal, a drastic evolution is seen when Li atoms enter sites situated within the central O plane, which causes a transformation from tin oxide to lithia. 
Following the modifications of the central plane, Sn atoms are pushed outwards and the majority of them share only three bonds with the neighbor O atoms. This should be compared to the four-fold coordinated sites within the pristine SnO crystal. The remaining Sn atoms located below the central plane are pushed even further outwards to sites where only a single bond to O remains. The inserted Li atoms occupy sites where they are connected to three nearest-neighbor  O atoms. All the O atoms keep their four-fold 
coordination and are located in tetrahedral sites with either two bonds to Li (1.9~\AA) and Sn (2.1~\AA), or one bond to Li (2.0~\AA) and three to Sn (2.2~\AA).
The expulsion of Sn atoms 
and
 the oxide transformation increases the volume of the structure by 23.14\%, as reported in Figure~\ref{fig:expansion}. It also modifies the symmetry of the supercell from  orthogonal to triclinic. Despite these significant changes it is important to realize that the layered and ordered structure persists. Furthermore, the segregated Sn atoms do not seem to assemble into nano-clusters, but rather to comply with the planar structure of the SnO matrix.

\subsubsection{LiOSn} 
\begin{figure}
\centering
\includegraphics[width=.5\textwidth]{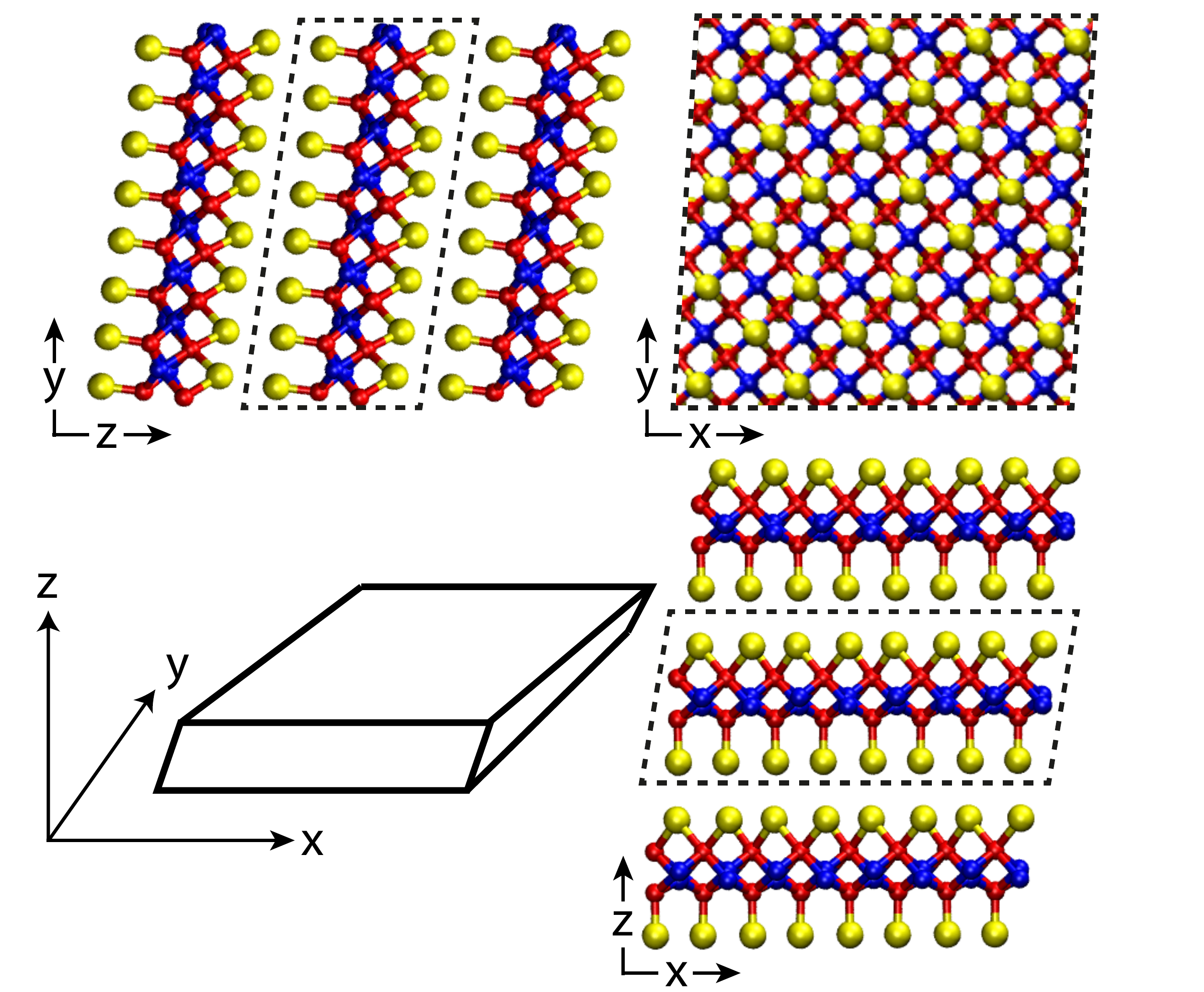}
\caption{LiOSn sample (color online). 
The color code and dashed box are defined as in Figure~\ref{fig:LiO_half_Sn}.
As for the Li$_{1/2}$OSn sample the (100) and (010) surfaces are identical but the structure has further evolved and the Sn atoms are now completely expelled from the oxide and reside in 'surface' planes on the forming lithia.  
}
\label{fig:LiOSn}
\end{figure}

As a next step LiOSn is generated by decorating all sites, within the Sn planes of the pristine SnO crystal with Li atoms.  During the structural relaxation and energy minimization the intrusion of Li atoms continues and all the O atoms in the central plane are replaced by Li.
The result of this transformation is depicted in Figure~\ref{fig:LiOSn} that represents a family of low energy configurations with some recombination in the interfacing Sn 'surface' layers. Another example is reported in the supplementary material.      
%
In the selected configuration the central plane is occupied by four-fold coordinated Li atoms that reside at tetrahedral sites with bonds to O of length 1.9~\AA. The O atoms are now located between the Li and Sn atoms in either a five-fold coordinated square pyramid with four bonds to Li in the base and a single bond to Sn in the apex (2.0~\AA) or a six-fold coordinated site in a truncated cube with four Li atoms in the the base and two Sn bonds of length 2.2~\AA\ in the top diagonal.
The Sn atoms are expelled from the central plane and are now residing as 'surface' planes on the emerging LiO slab. By carefully inspecting this structure it turns out that the Sn plane above and below the lithia are slightly different. The Sn atoms on the uppermost 'surface' reside at Li sites, whereas O sites are occupied in the plane below. Sn atoms originally belonging to two different neighbor layers are now located directly above each other in the (100) and (010) planes
but
the Sn atoms in the two layers are shifted in the (001) plane. Due to the chosen cut off when plotting bonds the Sn atoms at Li sites (above) appear to have two bonds, whereas the ones at O sites (below) only exhibit a single bond. 

As compared to the SnO reference crystal the volume expansion of LiOSn is 29.42\%, while it was 23.14\% for Li$_{1/2}$OSn. This moderate change indicates that the lithiation of SnO has reached a stable state, while the phase change (SnO $\rightarrow$ LiO) and a huge volume increase occurred during the previous step described in Section~\ref{sec:LiO_half_Sn}.
Again, the ordered and layered structure of the initial SnO sample remains. The sides of the supercell are skewed, as for the Li$_{1/2}$OSn sample. 

\begin{figure}
\centering
\includegraphics[width=.5\textwidth]{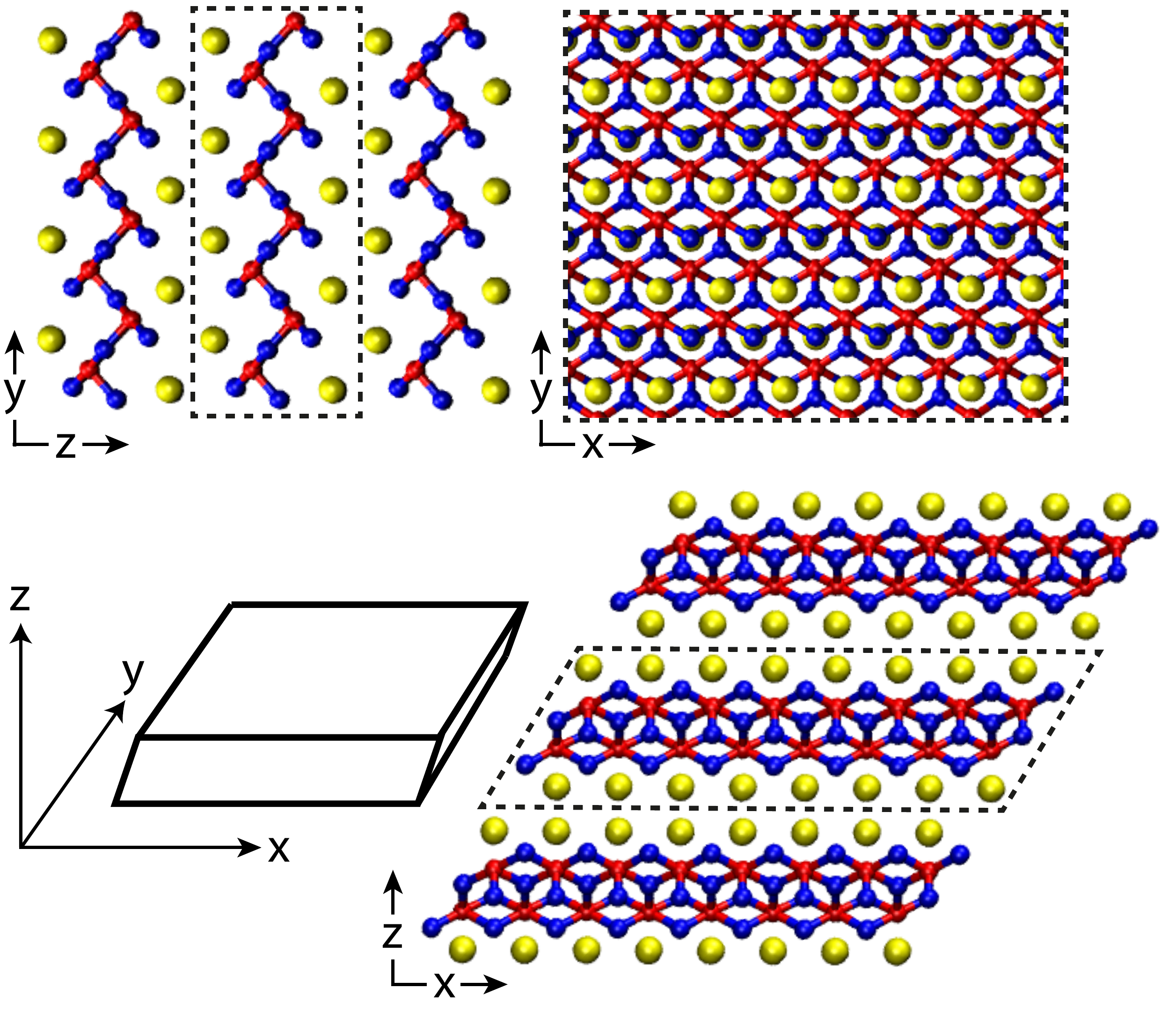}
\caption{Fully saturated Li$_2$OSn$^\text{L}$ (color online). 
The same color code as before is used. 
As the lithia saturates the super cell regains an orthogonal shape in the (001) plane, but is skewed along the Z-direction. The structure of the (100) and the (010) plane is no longer similar. 
However, the layered structure persists and Sn atoms remain as 'surface' planes on the lithia. All bonds connecting Sn atoms are now longer than the cut off distance, which was defined in Figure~\ref{fig:SnO}.
}
\label{fig:Li2OSn}
\end{figure}

\subsubsection{Li$_2$O$^\text{L}$Sn}
By inserting an additional plane of Li atoms into the LiOSn structure a fully saturated lithia is obtained. It is plotted in Figure~\ref{fig:Li2OSn}. All the O atoms have now a
six-fold coordination and reside at octahedral sites where they have five bonds of equal length (1.9~\AA) to the neighbor Li atoms and a single bond to Sn that is 2.3~\AA.
%
The Li atoms are either two-fold 
(interface planes) 
or three-fold coordinated (central plane). It is worthwhile noting that none of the Sn-Sn bonds are within the defined cutoff distance, as can be seen in Figure~\ref{fig:Li2OSn}. The 'surface' planes therefore appear as weakly bound to the enclosed lithia. Along the X-direction the spacing between the Sn atoms is 
3.0~\AA,
the distance along the Y-direction is
4.9~\AA,
while Sn atoms belonging to two neighbor layers are separated by  
3.3~\AA.
The shape of the supercell is monoclinic and only the (010) surface is skewed. However, the (100) and (010) planes are no longer equivalent.  As compared to the pristine SnO the volume expansion of the  Li$_2$O$^\text{L}$Sn is 58.59\%.
This number is consistent with the results for Li$_{1/2}$OSn and LiOSn since they are all composed of a central lithia that is surrounded by planes of Sn atoms.

%


\section{Discussion}
\label{sec:Discussion}
Our simulation results for the lithiation of SnO align very well with the experimental findings that Li behaves as an intercalated species at low concentrations~\cite{Sandu:2004gl}. 
A similar behavior of Li has been observed for SnO$_2$ in a computational study by Sensato {\it et al.}~\cite{Sensato:2012kk}
As the Li load level increases four-atom clusters form, as shown in Figure~\ref{fig:intercalated} (b).  Due to  their strong binding energy these clusters retard the emergence of continuous Li stripes. Superstructures made of lines of these low energy four-atom clusters
should
 appear instead.

%
At higher Li concentrations neighbor Li clusters are more likely to merge and initiate the oxide transformation through the proposed 'zipper' mechanism. 
Such an evolution is consistent with the stripe nucleation experimentally determined for SnO$_2$~\cite{Zhong:2011hs}.  
It cannot be demonstrated that the nucleation shown in Figures~\ref{fig:intercalated}~(c) and \ref{fig:LiO_half_Sn}, and the data reported in Reference~\citenum{Zhong:2011hs} result from the same physical / chemical process, but their similarity is striking. 
%
However, the
oxide transformation is likely to be a spontaneous process. 

As the lithia forms and saturates the expelled Sn atoms are equally distributed above and below the central O plane.  The outcome of this process is that the Sn atoms form 'surface' planes on the layers of the emerging lithia, which can be clearly seen  in Figures~\ref{fig:intercalated}~(c), \ref{fig:LiO_half_Sn}, \ref{fig:LiOSn} and \ref{fig:Li2OSn}. Since Sn agglomerates into planes outside the lithia, a distribution of various Sn-Sn, Sn-O and Sn-Li bonds occur. This is significantly different from a sample with Sn nano-clusters, which would be dominated by Sn-Sn bonds. Our structures with a wide bond distribution could explain the 'exotic' Sn bonds observed in experiments when a SnO sample is lithiated~\cite{Courtney:1997wl, Chouvin:2000kl, Sandu:2004gl}.

Another important aspect that validates our model is the good qualitative agreement with the volume expansion curve measured for Li$_\text{X}$OSn with respect to SnO and presented in Reference~\citenum{Ebner:2013bj}. As in the experimental data, three distinct regimes can be identified in our computational results shown in Figure~\ref{fig:expansion}:
(i) intercalation at low Li concentrations that results in a slow volume expansion,
(ii) oxide transformation when the four atom Li clusters start to merge, causing an abrupt volume change, and finally
(iii) lithia saturation when the number of Li atoms becomes similar to Sn, which slows down the volume expansion.
However, the determined Li load levels for the onset of the different regimes do not coincide with the experimental ones. This discrepancy can be explained by the fact that not all the initial SnO layers are lithiated at the same rate. We restricted our simulations to a single Li$_\text{X}$OSn layer and by imposing periodic boundary conditions along the X, Y, and Z directions, all the anode layers are assumed to be identical. In reality, as mentioned in Reference~\citenum{Sandu:2004gl}, parts of the electrode undergo an oxide transformation, while other parts are still able to absorb Li as an intercalated species. Hence, the simulated results should be interpreted as the upper boundary for the volume expansion during the lithiation process. For example Li$_{1/2}$OSn represents the end point for the oxide transformation in terms of volume expansion and marks the transition to the final regime (lithia saturation). The introduction of additional Li further saturates the oxide, leading to the formation of LiOSn and finally Li$_{2}$O$^\text{L}$Sn.      
%
%
%
%
%
During this phase the slope of the volume expansion is reduced, which agrees well with the experimental observations in Reference~\citenum{Ebner:2013bj}. Note that the volume expansion predicted when assuming totally segregated Li$_2$O and Sn crystals, the black square in Figure~\ref{fig:expansion}, underestimates the experimental value by 7.5~\%. In contrast the value predicted by our model exceeds the experimental data, but only by 3.1~\%.
As the distance between Sn atoms belonging to different layers decreases 
for increasing Li concentrations
the volume expansion appears to be caused by the transformation from tin 
oxide into lithia
and the formation of Li$_2$O$^\text{L}$Sn.

The layered Li$_2$O$^\text{L}$Sn structure is thermodynamically metastable as compared to a sample where Li$_2$O and Sn are segregated. Because the Sn atoms form bonds to Li$_2$O$^\text{L}$ and 
the lack of a Sn concentration gradient within the planes, the proposed structure should remain sufficiently stable as long as the electrode is operated under moderate loading and unloading conditions. 
The performance of a SnO anode has indeed been seen to retain its storage capacity if a reduced voltage range 
is applied~\cite{Courtney:1997br, Courtney:1997wl, Jeong:2013eh}.
%
If operated under extreme conditions an agglomeration of Sn into nano-clusters might occur for both SnO\cite{Courtney:1999du, Chao:2011dz} and SnO$_2$\cite{Zhang:2012kq}. But this is not the case that has been considered in this paper

A final point supporting the validity of our model is the electron conductivity: in a system with completely separated Li$_2$O and Sn the  electrons could not efficiently propagate from the metal contact to the active region of the anode or vice-versa, thus limiting the Li reduction or oxidation rate. In effect the electron conductivity of oxides is poor and the Sn clusters are not necessarily connected to each other to ensure electron transport. Since SnO has been demonstrated as a promising anode material, there must be well-conducting electron channels, as for example the 'surface' planes in Figures~\ref{fig:LiO_half_Sn}, \ref{fig:LiOSn}, and \ref{fig:Li2OSn}, which are made of Sn and directly connected to the lithia.

\section{Conclusions}
\label{sec:Conclusions}
The structure of Li$_{1/2}$OSn, LiOSn, and Li$_{2}$OSn has theoretically  been determined by lithiating a SnO sample. It has been found that during this process the volume expansion undergoes three distinct regimes consistent with the experimental observations of Reference~\citenum{Ebner:2013bj}: 
(i) slow volume changes when Li occupies intercalated sites; 
(ii) an increased volume expansion when Li reacts with O to form lithia;
(iii) reduced volume expansion when saturating the lithia.
The transformation from tin oxide to lithia is likely to be initiated form a LiO stripe resulting from a 'zipper' nucleation.

All three phases mentioned above keep the layered and ordered structure of the initial SnO crystal. The central Li$_\text{X}$O layer is embedded between Sn 'surface' planes. These planes allow us to explain the presence of 'exotic' Sn bonds when SnO is lithiated. The  final structure we studied, Li$_2$O$^\text{L}$Sn, is metastable as compared to a fully segregated and lower energy sample with Li$_2$O and Sn crystals.  In our case the Sn atoms residing between Li$_2$O$^\text{L}$ appear to be bound to these layers, which prevent Sn from segregating and forming nano-clusters if the battery is operated under moderate conditions. 

The fact that Sn remains in-between the Li$_2$O$^\text{L}$ layers challenges the common assumption that Sn agglomerates into clusters within a Li$_2$O matrix when SnO is lithiated. This key feature strongly influences the structural changes occurring when additional Li atoms are loaded into the SnO sample. This issue will be addressed in an upcoming paper.

\begin{acknowledgement}
The authors would like to acknowledge ERC for funding obtained from the starting grant (E-MOBILE) and CSCS for computational resources under project S503. They would also like to thank Dr. Martin Ebner at ETH Zurich for valuable inputs and fruitful discussions.
\end{acknowledgement}

\begin{suppinfo}
A complete list of the obtained lowest energy structures and a detailed description of how these have been prepared and tested are provided in the supporting material. Furthermore, the lowest energy structure for LiOSn is shown. 
\end{suppinfo}

\bibliography{SnO_lithiate}

\end{document}